\documentclass[journal]{IEEEtran}
\ifCLASSINFOpdf
\else
\fi
\usepackage[normalem]{ulem}
\usepackage{hyperref}
\usepackage{flushend}
\usepackage{csquotes}
\usepackage{lipsum}

\usepackage{soul}
\usepackage{graphicx}
\usepackage{amsmath}
\usepackage{amsthm}
\usepackage{float}
\usepackage{amssymb}
\usepackage{color}
\usepackage{algorithm}
\usepackage{multirow}
\usepackage{subfigure}
\usepackage{algpseudocode}

\usepackage{cite}
\useunder{\uline}{\ul}{}
\usepackage{url}
\usepackage{upgreek}
\graphicspath{{Figs/}} 

\theoremstyle{definition}

{
	\theoremstyle{plain}
	
}

\usepackage{scalerel}
\usepackage{tikz}
\usetikzlibrary{svg.path}

\definecolor{orcidlogocol}{HTML}{A6CE39}
\tikzset{
  orcidlogo/.pic={
    \fill[orcidlogocol] svg{M256,128c0,70.7-57.3,128-128,128C57.3,256,0,198.7,0,128C0,57.3,57.3,0,128,0C198.7,0,256,57.3,256,128z};
    \fill[white] svg{M86.3,186.2H70.9V79.1h15.4v48.4V186.2z}
                 svg{M108.9,79.1h41.6c39.6,0,57,28.3,57,53.6c0,27.5-21.5,53.6-56.8,53.6h-41.8V79.1z M124.3,172.4h24.5c34.9,0,42.9-26.5,42.9-39.7c0-21.5-13.7-39.7-43.7-39.7h-23.7V172.4z}
                 svg{M88.7,56.8c0,5.5-4.5,10.1-10.1,10.1c-5.6,0-10.1-4.6-10.1-10.1c0-5.6,4.5-10.1,10.1-10.1C84.2,46.7,88.7,51.3,88.7,56.8z};
  }
}

\newcommand\orcidicon[1]{\href{https://orcid.org/#1}{\mbox{\scalerel*{
\begin{tikzpicture}[yscale=-1,transform shape]
\pic{orcidlogo};
\end{tikzpicture}
}{|}}}}
\begin{document}

    \title{\huge Physically Guided Deep Unsupervised Inversion  for 1D Magnetotelluric Models}
    \author{Paul Goyes-Pe\~nafiel\textsuperscript{\orcidicon{0000-0003-3224-3747}} ~\IEEEmembership{Graduate Student Member,~IEEE,} Umair bin Waheed\textsuperscript{\orcidicon{0000-0002-5189-0694}}, ~\IEEEmembership{Member,~IEEE,} Henry Arguello\textsuperscript{\orcidicon{0000-0002-2202-253X}}, ~\IEEEmembership{Senior Member,~IEEE}
    
    \thanks{Paul ~Goyes-Pe\~nafiel and H.~Arguello are with the Department of Systems Engineering, Universidad Industrial de Santander, Bucaramanga 680002, Colombia (e-mail: goyes.yesid@gmail.com, henarfu@uis.edu.co)}
    \thanks{Paul ~Goyes-Pe\~nafiel and H.~Arguello acknowledge the support provided by the Universidad Industrial de Santander under Project 3925.}
    \thanks{Umair bin Waheed is with the Department of Geosciences, King Fahd
University of Petroleum and Minerals, Dhahran 31261, Saudi Arabia (e-mail:
umair.waheed@kfupm.edu.sa)}
    \thanks{Umair bin Waheed acknowledges the support provided by the College of Petroleum Engineering and Geosciences (CPG) at KFUPM through Competitive Research Grant No. CPG21103.} \vspace*{-1cm}
    }
    
    \markboth{}%
    {Shell \MakeLowercase{\textit{et al.}}: Bare Demo of IEEEtran.cls for IEEE Journals}
    \maketitle

\begin{abstract}
The global demand for unconventional energy sources such as geothermal energy and white hydrogen requires new exploration techniques for precise subsurface structure characterization and potential reservoir identification. The Magnetotelluric (MT) method is crucial for these tasks, providing critical information on the distribution of subsurface electrical resistivity at depths ranging from hundreds to thousands of meters. However, traditional iterative algorithm-based inversion methods require the adjustment of multiple parameters, demanding time-consuming and exhaustive tuning processes to achieve proper cost function minimization. Recent advances have incorporated deep learning algorithms for MT inversion, primarily based on supervised learning, and large labeled datasets are needed for training. This work utilizes TensorFlow operations to create a differentiable forward MT operator, leveraging its automatic differentiation capability. Moreover, instead of solving for the subsurface model directly, as classical algorithms perform, this paper presents a new deep unsupervised inversion algorithm guided by physics to estimate 1D MT models. Instead of using datasets with the observed data and their respective model as labels during training, our method employs a differentiable modeling operator that physically guides the cost function minimization, making the proposed method solely dependent on observed data. Therefore, the optimization algorithm updates the network weights to minimize the data misfit. We test the proposed method with field and synthetic data at different acquisition frequencies, demonstrating that the resistivity models obtained are more accurate than those calculated using other techniques. Our implementation is available at \textcolor{magenta}{\url{https://github.com/PAULGOYES/MT_guided1DInversion.git}}.
\end{abstract}


\section{Introduction}\label{introduction}

\IEEEPARstart{M}AGNETOTELLURIC  (MT) surveys have emerged as a powerful geophysical technique for probing the electrical conductivity structure of the Earth's subsurface. With applications ranging from mineral exploration and unconventional energy resources to groundwater mapping and volcano monitoring, MT offers unparalleled insights into subsurface geological features \cite{angeles2013,Zhdanov2011}. MT equipment measures natural variations of the Earth's electromagnetic fields over a frequency range. By analyzing the impedance tensor related to electric and magnetic fields, MT surveys provide valuable information about electrical conductivity distribution in the subsurface \cite{Guo2011, Feng2022, jupp1977}. However, accurately interpreting MT data and deriving robust subsurface resistivity models remain challenging tasks, underscoring the importance of inversion techniques to mitigate uncertainties \cite{Guo2011}.

In the realm of MT inversion methods, two predominant approaches have traditionally prevailed: iterative solutions and approximate methods such as Gauss-Newton, inexact Newton \cite{grayver2013}, or conjugate gradient algorithms \cite{newman2000, Candansayar2008}. These methods often necessitate prior information about the subsurface model through regularization techniques like initial models or sparsity constraints, and they typically require tuning numerous parameters \cite{Kang2017}. Recent efforts have shifted towards leveraging deep learning techniques to overcome these limitations. These approaches employ supervised learning to extract features and establish relationships between observations and subsurface properties from training datasets. Nonetheless, they often encounter challenges associated with the dataset size and the fixed structure of neural networks, which may affect generalization because the models used during training do not represent the domain of data acquired in the field. To reduce dataset dependency, \cite{Liu2022} have implemented physics-based regularization operators to lessen reliance solely on the training dataset.

In the current state of the art regarding the estimation of subsurface conductivity models from magnetotelluric measurements, convolutional \cite{Liao2022}, residual, and autoencoder neural networks are predominantly utilized \cite{RahmaniJevinani2024}. Additionally, efforts have been made to expedite the generation of reliable models and their measurements using deep learning to approximate forward modeling and obtain apparent resistivity and phase angle curves \cite{Deng2023a}. However, the magnetotelluric response of the subsurface is sensitive to the resistive anomalies that could cause significant variations in deep learning-based inversion models, as they rely on the statistical distribution of thicknesses and resistivities used to create the dataset \cite{Rung-Arunwan2022}.

In this work, we introduce a new approach to MT data inversion that overcomes the limitations of traditional deep learning methods. We have created a differentiable tensor-based forward modeling operator to eliminate the need for external data during inversion. This operator computes electromagnetic impedance, allowing integration into an unsupervised learning framework where a neural network approximates the inverse operator. Notably, the cost function relies solely on measured data, speeding up the computation process and avoiding the use of external data. Our method represents a significant advancement in MT inversion, providing a promising path to more reliable and efficient subsurface imaging.

\section{Methodology}
\subsection{Forward MT operator}
\label{Forward_opt}

MT involves measuring fluctuations in the natural electric ($\mathbf{E}$) and magnetic ($\mathbf{B}$) fields in orthogonal directions (i.e., $xy$ or $yx$) on the Earth's surface in the frequency domain \cite{Zhdanov2017}. In the 1D approximation, the relationship between these fields is described by a $2 \times 2$ matrix known as the impedance tensor: 
$$ Z = \begin{pmatrix}
0 & Z_{xy} \\
-Z_{xy} &  0\\
\end{pmatrix} . $$

It should be noted that impedance is a complex number that depends solely on conductivity and frequency within a homogeneous layer. For layered earth, the impedance component $Z_{xy}$ can be expressed as the total contribution of every single layer at different depths $z$ as is described in the following equation:
\begin{equation}
\label{eq:impedance}
    Z_{xy}(z) = E_x(z) /H_y(z) = \frac{\omega \mu_0}{k_j},
\end{equation}
where $\mu_0$ is the vacuum magnetic permeability and $k_j = \sqrt{i\omega \mu_0 \sigma_j}$ is the wave number at some angular frequency $\omega$ within $j$-th earth layer with conductivity $\sigma_j$. Note that \eqref{eq:impedance} defines the electric and magnetic fields from the solution of a plane wave propagation in layered earth such as $E_x(z) = E_{x_j}^{+} e^{ik_j z}$ and $H_y(z) = \frac{k_j}{\omega \mu_0} E_{x_j}^{+} e^{ik_j z}$. By convention, $z_{N-1}+0$ is the depth to the top of the layer $N$ such as $z_{N-1}+0 \leq z \leq \infty$. Therefore, the wave impedance from layer $j=N-1$ to $j=1$ is given as:

\begin{equation}
\label{eq:Zfull}
\resizebox{.91\hsize}{!}{$ Z_{xy}(z_{j-1}+0)= - \frac{\omega \mu_0}{k_j}\operatorname{coth} \Bigl \{ ik_j h_j  - \operatorname{coth}^{-1} \Bigl [ \frac{k_j}{\omega \mu_0} Z_{xy}(z_j+0) \Bigr ] \Bigr \}$.}
\end{equation}

It is worth noting that \eqref{eq:Zfull} is recursive. Following \cite{Zhdanov2017}, we adopt the next expression to calculate the impedance at the Earth's surface (i.e., $z=0$): 

\begin{equation}
\label{eq:impedance_rn}
    Z_{xy}(+0)= \frac{\omega \mu_0}{k_1} R_{j},
\end{equation}
where $R_j$ is the layered-earth correction factor for the plane wave impedance for the layered medium with thicknesses $h_j$.

We formulate forward modeling as an operator $\pmb{\mathcal{F}}$ parameterized by angular frequency $\{ \omega \}_{i=1}^{J}$ and the model $\mathbf{m}=  \left \{ \{ h_{j} \}_{j=1}^{N-1};  \{ \sigma_{j} \}_{j=1}^{N}    \right \}$ with thicknesses $h_j$ and conductivities  $\sigma_j$. Note that the layer $N$ represents the half-space with infinity thickness and $J$ is the number of frequencies. Hence, our operator obtains the electromagnetic impedance response $\mathbf{d} \in \mathbb{C}^J$ at the surface using \eqref{eq:impedance_rn} and is described by the following expression:

\begin{equation}
    \label{eq:forward}
    \mathbf{d} = Z_{xy}(+0) = \pmb{\mathcal{F}}[\mathbf{m}].
\end{equation}

We implemented operator $\pmb{\mathcal{F}}$  using TensorFlow operations to exploit its capability for automatic differentiation \cite{tensorflow2015-whitepaper}. This allows for integrating our operator with neural networks as a single computational graph. As a result, the operator is differentiable, allowing computing gradients and use it in backpropagation. The code implementation is available in the project repository.

\subsection{Neural inverse operator}

We approximate the inverse operator by a neural network $\mathcal{I}_{\mathbf{\Theta}}$ with trainable parameters $\Theta$ which satisfies the following equivalence: 

\begin{equation}
\label{eq:inverseopt}
    \mathcal{I}_{\mathbf{\Theta}}(\mathbf{d}) \approx \mathbf{m}.
\end{equation}

\begin{figure}[ht]
    \centering
    \includegraphics[width=1\columnwidth]{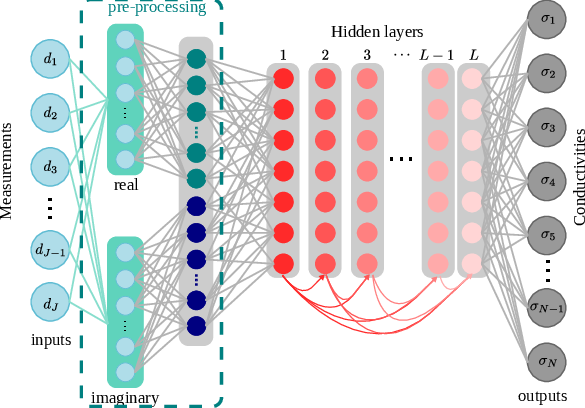}
    \vspace*{-6mm}
    \caption{The neural network $\mathcal{I}_{\mathbf{\Theta}}$ has inputs given by $\mathbf{d}=[d_1, d_2, d_3, \cdots, d_J]$ and outputs with the conductivities of the model $\pmb{\sigma}=[\sigma_1, \sigma_2, \sigma_3, \cdots, \sigma_N]$. The number of neurons is constant within the $L$ hidden layers.}
    \label{fig:network}
\end{figure}

The proposed method features a neural network $\mathcal{I}_{\mathbf{\Theta}}$ with a pre-processing module that extracts and concatenates the real and imaginary parts of $\mathbf{d}$ to form the initial layer. This layer connects to hidden layers, which consist exclusively of fully connected layers. These layers produce cumulative outputs by adding previous layers' outputs to the current output layer, establishing shortcut connections, as shown in Fig.~\ref{fig:network}. Shortcut connections facilitate direct gradient flow through the network and preserve gradient magnitude during backpropagation \cite{Feng2019AnDenseNet}. 

\begin{algorithm}
\caption{Fully Connected Network with Additive Layers}
\label{alg:network}
\begin{algorithmic}[1]
\Require $\mathbf{d}$ \Comment{Input data with dimension $J$}
\Require $L$ \Comment{Number of hidden layers}
\Ensure $\pmb{\sigma}$ \Comment{Output data with dimension $N$}

\State $\mathbf{h}_0 \gets \{\Re(\mathbf{d}), \Im(\mathbf{d}) \}$ \Comment{Extract and concatenate real and imaginary parts}
\State $\mathbf{a}_0 \gets \text{ReLU}(\mathbf{W}_0 \mathbf{h}_0 + \mathbf{b}_0)$ \Comment{Output of the first hidden layer}
\State $\mathbf{s}_0 \gets \mathbf{a}_0$ \Comment{Initial sum of hidden layers output}

\For {$i = 1$ to $L$} \Comment{For each hidden layer}
    \State $\mathbf{h}_i \gets \text{ReLU}(\mathbf{W}_i \mathbf{s}_{i-1} + \mathbf{b}_i)$ 
    \State $\mathbf{s}_i \gets \mathbf{s}_{i-1} + \mathbf{h}_i$ \Comment{Add to previous sum}
\EndFor

\State $\pmb{\sigma} \gets \text{Sigmoid}(\mathbf{W}_\text{o} \mathbf{s}_{L} + \mathbf{b}_\text{o})$ \Comment{Compute final output with sigmoid activation}

\State \Return $\pmb{\sigma}$
\end{algorithmic}
\end{algorithm}

The detailed steps of the proposed neural inverse operator are shown in Algorithm~\eqref{alg:network} where the trainable parameters $\mathbf{\Theta}$ feature a fully connected structure with multiple hidden layers with matrix weights $\mathbf{W}$ and bias $\mathbf{b}$, along with an input layer consisting of $J$ neurons and an output layer of $N$ neurons, representing the number of frequencies and conductivities, respectively.

\subsection{Physically guided unsupervised inversion} 

We adopted the state-of-the-art formulation for the geophysical optimization problem in terms of data misfit $\Phi_d$ and model norm $\Phi_m$ \cite{Liu2024,Feng2022,Kang2017} given by the Tikhonov regularization as follows:

\begin{equation}
\label{eq:tikhonov}
    \Phi(\mathbf{m})=\Phi_d (\mathbf{m}) +\lambda \Phi_m(\mathbf{m}).
\end{equation}

Optimizing \eqref{eq:tikhonov} involves finding the optimal model $\mathbf{m}$. We rewrite the optimization problem, substituting  \eqref{eq:inverseopt} into \eqref{eq:tikhonov}. Therefore, the proposed objective function \eqref{eq:costfnc} is minimized when finding the optimal network parameters $\mathbf{\Theta}$ 

\begin{equation}
\label{eq:costfnc}
    \Phi(\mathbf{\Theta})=\Phi_d(\mathbf{\Theta})+\lambda \Phi_m(\mathbf{\Theta}), 
\end{equation}
where $\lambda$ is a trade-off parameter weighting the importance between data misfit and model regularization. We included our inverse operator $\mathcal{I}_{\mathbf{\Theta}}$ to define $\Phi_d(\mathbf{\Theta})$ and  $\Phi_m(\mathbf{\Theta})$ given by the equations \eqref{eq:datamisfit} and \eqref{eq:modelmisfit}, respectively.

\begin{equation}
\label{eq:datamisfit}
    \Phi_d(\mathbf{\Theta}) = \frac{1}{2} \left \| \pmb{\varepsilon}_d(\pmb{\mathcal{F}}[\mathcal{I}_{\mathbf{\Theta}}(\mathbf{d}_{\text{obs}})] - \mathbf{d}_{\text{obs}}) \right \|_2, 
\end{equation}

\begin{equation}
\label{eq:modelmisfit}
    \Phi_m(\mathbf{\Theta}) = \frac{1}{2} \left \| \mathbf{m}_{\text{ref}}- \mathcal{I}_{\mathbf{\Theta}}(\mathbf{d}_{\text{obs}}) \right \|_2, 
\end{equation}

where $\mathbf{d}_{\text{obs}}$ is the measured impedance response data. $\pmb{\varepsilon}_d = \mathrm{diag}(1/\epsilon_i)$ with $\epsilon_i$ the standard deviation of the $i$-th measurement. $\mathbf{m}_{\text{ref}}$ is a reference model. 

The inverse neural network $\mathcal{I}_{\mathbf{\Theta}}$ is trained unsupervised by minimizing \eqref{eq:costfnc}, guided by the forward MT operator $\pmb{\mathcal{F}}$. Therefore, our inversion algorithm relies solely on the measured data rather than datasets typically used in supervised deep learning.  

\begin{algorithm}
\caption{Deep unsupervised MT inversion}
\label{alg:inv}
\begin{algorithmic}[1]
\Require $\pmb{\mathcal{F}}$, $K$,  $\mathbf{d}_{\text{obs}}$, $\eta$ learning rate, $\mathbf{m}_{\text{ref}}$ (optional)
\Ensure Inverted resistivity model $\mathbf{m}$
\State  \textbf{Initialize} $\mathcal{I}_{\mathbf{\Theta}}$ with $\mathbf{\Theta}_1$ glorot uniform distribution
\For{$i = 1,\dots,K$} 

\State Calculate the objective function for the current network parameters $$\Phi(\mathbf{\Theta}_i)=\Phi_d(\mathbf{\Theta}_i)+\lambda \Phi_m(\mathbf{\Theta}_i)$$
\State Minimize \eqref{eq:costfnc} using an optimizer and update the trainable parameters $$\mathbf{\Theta}_{i+1}  \gets \mathbf{\Theta}_{i} - \eta \nabla_{\mathbf{\Theta}} \Phi(\mathbf{\Theta}_i)  $$
\EndFor

\State $\mathbf{m} \gets \mathcal{I}_{\mathbf{\Theta}_{K}}(\mathbf{d}_{\text{obs}})$ \Comment{Predict inverted model using the parameters from the last iteration $K$. }

\State \textbf{return:} $\mathbf{m}$  
\end{algorithmic}
\end{algorithm}

Algorithm \ref{alg:inv} outlines step-by-step the proposed methodology, where it is evident that the only mandatory parameters for conducting the inversion are the number of epochs $K$ and the learning rate $\eta$. However, it is also common practice to incorporate prior information, such as an initial reference model $\mathbf{m}_{\text{ref}}$, upon which additional subsurface information can be included.  Step 1 initializes the trainable parameters $\mathbf{\Theta}$ using a glorot uniform distribution \cite{pmlr-v9-glorot10a}. In step 3, the cost function value is calculated for each epoch. Subsequently, in step 4, a backpropagation algorithm is applied to update the trainable parameters and minimize \eqref{eq:costfnc}. It is important to highlight that our proposed method does not require an initial model, unlike gradient-based or Gauss-Newton methods \cite{grayver2013,newman2000, Candansayar2008}. Finally, the observations are evaluated in the neural network, and the subsurface model is obtained.

\section{Results and discussion}

\textbf{Implementation details:}
We implemented a fully connected neural network with $L=5$ hidden layers of 256 neurons with ReLU activations. The number of trainable parameters was 342,292. The number of neurons in the input and output layers depends on the number of frequencies and thicknesses for each experiment. We used a constant learning rate $\eta= 10^{-3}$ and an AdamW optimizer \cite{Adamw} for all experiments. The stopping criterion involved monitoring the cost function to ensure no improvement after 10 consecutive epochs. Note that the output layer uses the sigmoid activation function. Therefore, given the values $\rho_{\text{min}}$ and $\rho_{\text{max}}$, the network output is denormalized to extract the inverted resistivity:

\begin{equation}
    \rho_{\text{inverted}}= (1/\mathcal{I}_{\mathbf{\Theta}}(\mathbf{d}))*(\rho_{\text{max}}-\rho_{\text{min}}) + \rho_{\text{min}}
\end{equation}

For Experiments I and II we used  $\rho_{\text{min}}=10^0$ $\Omega$m and $\rho_{\text{max}}=10^3$ $\Omega$m.

\subsection{Experiment I:} We evaluated the proposed method on a simple three-layer model with theoretical thicknesses of 2500 m for each layer and the depth to the top of the half-space layer at 5000 m. The MT acquisition was simulated with a frequency range from 10$^{-3}$ Hz to 10$^{2}$ Hz. For inversion, we used a ten-layer model with a thickness of 555 m for each layer. In this experiment, we compared the proposed method with the solution of \eqref{eq:tikhonov} i.e., $\ell_2$--norm, which was initialized with a 100 $\Omega$m constant model. Note that for both cases, we used the same forward MT operator. We used a standard deviation $\epsilon_i = 1$ in this experiment.

\begin{figure}[ht]
    \centering
    \includegraphics[width=.9\columnwidth]{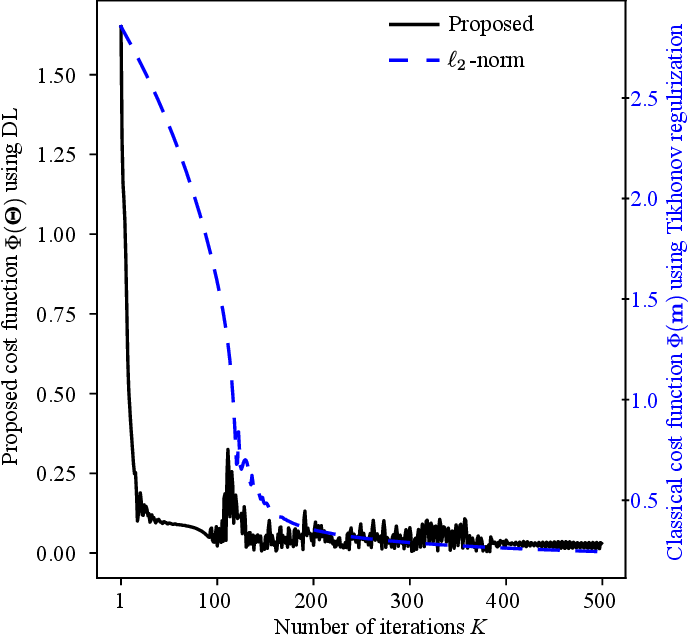}

    \caption{Data misfit curves over 500 iterations for the proposed method (left) from Eq. \eqref{eq:datamisfit} in black line and the solution based solely on the $\ell_2$--norm with Tikhonov regularization (right) from  Eq. \eqref{eq:tikhonov} in blue dashed line.}
    \label{fig:exp1loss}
\end{figure}

\begin{figure}[ht]
    \centering
    \includegraphics[width=1\columnwidth]{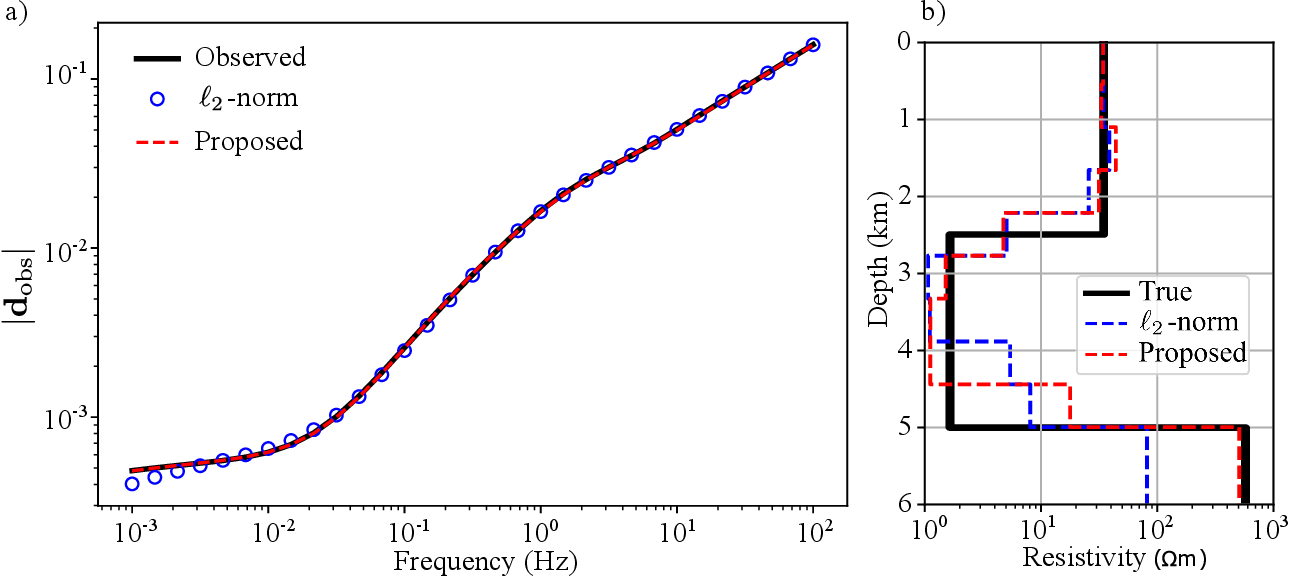}
    \vspace*{-6mm}
    \caption{Visual comparison of inversion results with the proposed method and the $\ell_2$-norm: a) in the impedance response and b) in the 1D resistivity model.}
    \label{fig:exp1inversion}
\end{figure}

Fig.~\ref{fig:exp1loss} shows the convergence curve for the cost function minimization. The $\ell_2$--norm exhibits a smooth curve stabilizing after iteration 200, reaching a minimum error of approximately 0.35. On the other hand, the proposed method, which includes the inverse neural operator, displays a steep curve in the first five iterations and stabilizes after iteration 125 with small fluctuations around approximately 0.15. This indicates that the proposed method achieves faster convergence of the data misfit for the inverse problem. Although both curves converged, Fig.~\ref{fig:exp1inversion}a shows that the $\ell_2$--norm formulation exhibits a mismatch, particularly at the lowest frequencies, whereas the proposed method fits all observations across all frequencies. This translates to differences in the inverted models in Fig.~\ref{fig:exp1inversion}b, mainly in the last layer (i.e., at a depth of 5 km), where the proposed method shows a good fit. The computation times are 85.543 s and 81.346 s for the proposed and the $\ell_2$--norm formulation, respectively.

\subsection{Experiment II:} We tested the inversion algorithm's capability on data where subsurface layers have variable thicknesses to simulate a realistic subsurface scenario. Our results were compared with the SimPEG inversion framework \cite{Kang2017, Heagy2017}, which incorporates modeling operators based on Maxwell's solutions to simulate electromagnetic fields with plane waves and a Gauss-Newton approach that employs the conjugate gradient solver \cite{Cockett2015}. SimPEG was initialized with a homogeneous resistivity model of 100 $\Omega$m. We used  20 layers with a maximum depth of 15,473 meters for both inversion methods. The proposed method used $K=100$ epochs. In this experiment, we found the best parameter $\lambda=0.0001$ from Eq.~\eqref{eq:costfnc}. The standard deviation in this experiment is generated from a Gaussian distribution, scaled by a 1\% relative error, and combined with the original data to simulate realistic measurement errors.

\begin{figure}[ht]
    \centering
    \includegraphics[width=1\columnwidth]{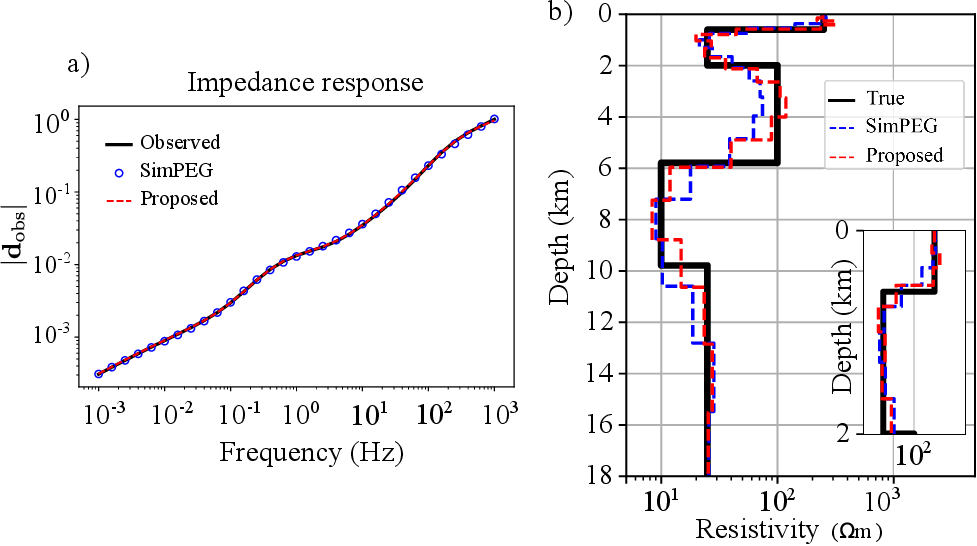}
    \vspace*{-6mm}
    \caption{a) Impedance response where both methods fit the measurements with high accuracy. b) The resistivity model shows that the proposed method performs better than SimPEG in delineating the first four layers, especially in areas with sharp resistivity changes. In contrast, SimPEG exhibits smooth resistivity transitions between consecutive layers.}
    \label{fig:exp2}
\end{figure}

Fig.~\ref{fig:exp2}a shows the impedance response for the inversions, demonstrating a high fit to the observed data for both SimPEG and the proposed method, with a normalized RMSE below 1\%. Fig.~\ref{fig:exp2}b shows that the proposed method better recovers resistivity changes, especially for the 2 km to 6 km depth layer. It is important to note that although the actual model has a maximum depth of 10 km, both methods exhibit variations in the theoretical resistivity for the final layer, as expected in magnetotelluric methods. It is important to note that the SimPEG algorithm is robust and requires fewer iterations. However, it necessitates tuning several parameters, including the number of iterations, stopping directives, smallness, smoothness, beta, beta cooling, etc. In contrast, our method requires only the learning rate and the number of iterations, which are easier to tune.

\vspace*{-5mm}
\subsection{Experiment III:} We tested the proposed method using field data from Earthscope and IRIS \cite{USGS}. The observed data were acquired at Yellowstone, WY, USA. The impedance response comprises 37 frequencies from $9.7\times 10^{-4}$ Hz to $2.5 \times 10^2$ Hz. In addition to SimPEG, we also compare the proposed method with the well-known software ZondMT1D based on least square inversion \cite{Patel2020}. The inversion was configured for all methods with 31 layers and a maximum and minimum resistivity of 1000 $\Omega$m and 10 $\Omega$m, respectively. The standard deviation in this experiment was modeled according to the record in the impedance files (.EDI) as a Gaussian distribution scaled by a 5\% relative error. Additionally, the maximum inversion depth was set to 59 km. The proposed method used $K=100$ epochs.

\begin{figure}[ht]
    \centering
    \includegraphics[width=1\columnwidth]{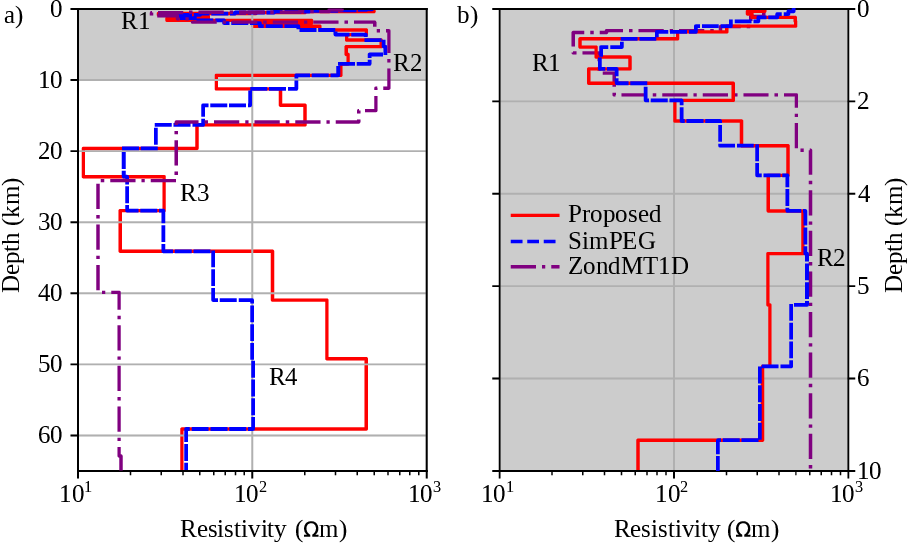}
    \vspace*{-6mm}
    \caption{a) 1D magnetotelluric inverted models with the proposed method compared with SimPEG \cite{Kang2017} and ZondMT1D \cite{Patel2020} inversions. b) Zoom between 0-10 km, highlighting the shallow resistivity anomalies R1 and R2. }
    \label{fig:exp3}
\end{figure}

Fig.~\ref{fig:exp3}a shows the inverted models, revealing similarities in the subsurface structure, where four resistivity anomalies can be identified. R1 and R2 are zoomed in Fig.~\ref{fig:exp3}b, located at depths of 1.5 km and 11 km, and were successfully detected by all inversion methods. The deep high-resistivity anomalies, R3 and R4, were only identified by the proposed method and SimPEG.

\begin{figure}[ht]
    \centering
    \includegraphics[width=1\columnwidth]{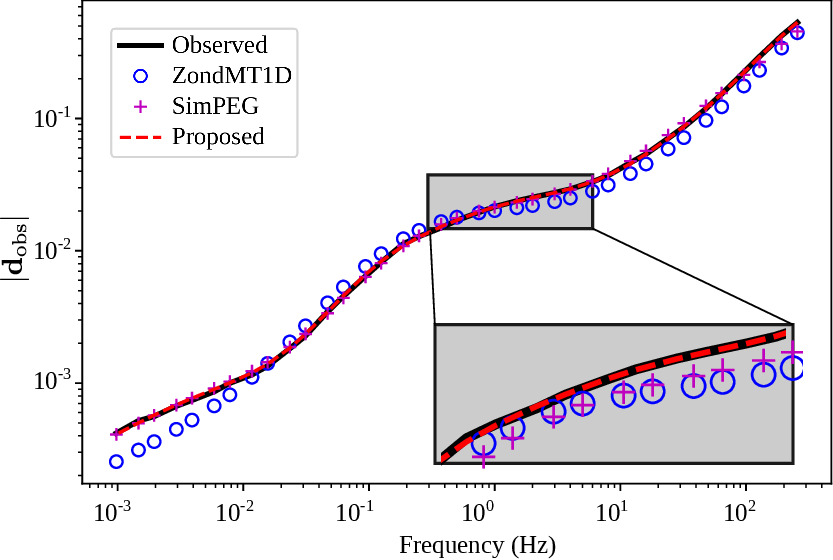}
    \vspace*{-6mm}
    \caption{Impedance response from the inverted models by the proposed method compared with SimPEG \cite{Kang2017} and ZondMT1D \cite{Patel2020} inversions. }
    \label{fig:exp3-2}
\end{figure}

Fig.~\ref{fig:exp3-2} shows the impedance response for the inversions using the proposed method compared to SimPEG and ZondMT1D. It is important to note that ZondMT1D has the poorest fit for the lowest frequencies (i.e., the deepest part of the model). This is evident in its failure to detect the R4 resistivity anomaly. While the fit is good for the remaining frequencies, the proposed method shows the best alignment, as highlighted in the zoomed-in area of Fig.~\ref{fig:exp3-2}. Quantitatively, the lowest normalized RMSE is achieved with the proposed method at 1.85\%, compared to ZondMT and SimPEG with 12.84\% and 2.84\%, respectively.

\vspace*{-2.5mm}
\section{conclusions}

In this work, we proposed approximating the magnetotelluric inverse operator using a neural network with additive layers. This approach enables the derivation of 1D resistivity models from impedance data. Our scheme is unsupervised and relies solely on a differentiable modeling operator constructed with tensor operations, allowing its integration as a forward operator in the neural network's output layer. We compared our method with SimPEG and ZondMT1D using synthetic scenarios and field data. The results demonstrated that our approach better fits the observed data and produces subsurface models with structures similar to those obtained by the compared methods. Quantitatively, our method achieved the best performance metrics.



\ifCLASSOPTIONcaptionsoff
  \newpage
\fi
    
    \bibliography{bib/refs}

\begin{thebibliography}{10}

\bibitem{millane1990phase}
R.~P. Millane, ``Phase retrieval in crystallography and optics,'' {\em JOSA A},
  vol.~7, no.~3, pp.~394--411, 1990.

\bibitem{pinilla2018coded}
S.~Pinilla, J.~Poveda, and H.~Arguello, ``Coded diffraction system in x-ray
  crystallography using a boolean phase coded aperture approximation,'' {\em
  Optics Communications}, vol.~410, pp.~707--716, 2018.

\bibitem{pinilla2018coded1}
S.~Pinilla, H.~Garc{\'\i}a, L.~D{\'\i}az, J.~Poveda, and H.~Arguello, ``Coded
  aperture design for solving the phase retrieval problem in x-ray
  crystallography,'' {\em Journal of Computational and Applied Mathematics},
  2018.

\bibitem{poon2014introduction}
T.-C. Poon and J.-P. Liu, {\em Introduction to modern digital holography: with
  MATLAB}.
\newblock Cambridge University Press, 2014.

\bibitem{shimano2018lensless}
T.~Shimano, Y.~Nakamura, K.~Tajima, M.~Sao, and T.~Hoshizawa, ``Lensless
  light-field imaging with fresnel zone aperture: quasi-coherent coding,'' {\em
  Applied optics}, vol.~57, no.~11, pp.~2841--2850, 2018.

\bibitem{fienup1987phase}
C.~Fienup and J.~Dainty, ``Phase retrieval and image reconstruction for
  astronomy,'' {\em Image Recovery: Theory and Application}, pp.~231--275,
  1987.

\bibitem{durig1986near}
U.~D{\"u}rig, D.~W. Pohl, and F.~Rohner, ``Near-field optical-scanning
  microscopy,'' {\em Journal of applied physics}, vol.~59, no.~10,
  pp.~3318--3327, 1986.

\bibitem{shechtman2015phase}
Y.~Shechtman, Y.~C. Eldar, O.~Cohen, H.~N. Chapman, J.~Miao, and M.~Segev,
  ``Phase retrieval with application to optical imaging: a contemporary
  overview,'' {\em IEEE signal processing magazine}, vol.~32, no.~3,
  pp.~87--109, 2015.

\bibitem{loewen2018diffraction}
E.~G. Loewen and E.~Popov, {\em Diffraction gratings and applications}.
\newblock CRC Press, 2018.

\bibitem{rodenburg2008ptychography}
J.~Rodenburg, ``Ptychography and related diffractive imaging methods,'' {\em
  Advances in Imaging and Electron Physics}, vol.~150, pp.~87--184, 2008.

\bibitem{thibault2009probe}
P.~Thibault, M.~Dierolf, O.~Bunk, A.~Menzel, and F.~Pfeiffer, ``Probe retrieval
  in ptychographic coherent diffractive imaging,'' {\em Ultramicroscopy},
  vol.~109, no.~4, pp.~338--343, 2009.

\bibitem{bendory2017fourier}
T.~Bendory, R.~Beinert, and Y.~C. Eldar, ``Fourier phase retrieval: Uniqueness
  and algorithms,'' in {\em Compressed Sensing and its Applications},
  pp.~55--91, Springer, 2017.

\bibitem{candes2013phaselift}
E.~J. Candes, T.~Strohmer, and V.~Voroninski, ``Phaselift: Exact and stable
  signal recovery from magnitude measurements via convex programming,'' {\em
  Communications on Pure and Applied Mathematics}, vol.~66, no.~8,
  pp.~1241--1274, 2013.

\bibitem{candes2015phase}
E.~J. Candes, X.~Li, and M.~Soltanolkotabi, ``Phase retrieval from coded
  diffraction patterns,'' {\em Applied and Computational Harmonic Analysis},
  vol.~39, no.~2, pp.~277--299, 2015.

\bibitem{chen2015solving}
Y.~Chen and E.~Candes, ``Solving random quadratic systems of equations is
  nearly as easy as solving linear systems,'' in {\em Advances in Neural
  Information Processing Systems}, pp.~739--747, 2015.

\bibitem{pinilla2018phase}
S.~Pinilla, J.~Bacca, and H.~Arguello, ``Phase retrieval algorithm via
  nonconvex minimization using a smoothing function,'' {\em IEEE Transactions
  on Signal Processing}, vol.~66, no.~17, pp.~4574--4584, 2018.

\bibitem{wang2018solving}
G.~Wang, G.~B. Giannakis, and Y.~C. Eldar, ``Solving systems of random
  quadratic equations via truncated amplitude flow,'' {\em IEEE Transactions on
  Information Theory}, vol.~64, no.~2, pp.~773--794, 2018.

\bibitem{wang2018phase}
G.~Wang, G.~B. Giannakis, Y.~Saad, and J.~Chen, ``Phase retrieval via
  reweighted amplitude flow,'' {\em IEEE Transactions on Signal Processing},
  vol.~66, no.~11, pp.~2818--2833, 2018.

\bibitem{candWir}
E.~J. Candes, X.~Li, and M.~Soltanolkotabi, ``Phase retrieval via wirtinger
  flow: Theory and algorithms,'' {\em IEEE Transactions on Information Theory},
  vol.~61, no.~4, pp.~1985--2007, 2015.

\bibitem{zhang2017nonconvex}
H.~Zhang, Y.~Zhou, Y.~Liang, and Y.~Chi, ``A nonconvex approach for phase
  retrieval: Reshaped wirtinger flow and incremental algorithms,'' {\em The
  Journal of Machine Learning Research}, vol.~18, no.~1, pp.~5164--5198, 2017.

\bibitem{chi2018nonconvex}
Y.~Chi, Y.~M. Lu, and Y.~Chen, ``Nonconvex optimization meets low-rank matrix
  factorization: An overview,'' {\em arXiv preprint arXiv:1809.09573}, 2018.

\bibitem{pinilla2019frequency}
S.~Pinilla, T.~Bendory, Y.~C. Eldar, and H.~Arguello, ``Frequency-resolved
  optical gating recovery via smoothing gradient,'' {\em arXiv preprint
  arXiv:1902.09447}, 2019.

\bibitem{pinilla2019exact}
S.~Pinilla, J.~Bacca, C.~Vargas, J.~C. Poveda-Jaramillo, and H.~Arguello,
  ``Exact crystalline structure recovery in x-ray crystallography from coded
  diffraction patterns,'' {\em arXiv preprint arXiv:1907.03547}, 2019.

\bibitem{bacca2018super}
J.~Bacca, S.~Pinilla, D.~Molina, A.~Camacho, and H.~Arguello,
  ``Super-resolution phase retrieval algorithm using a smoothing function,'' in
  {\em Mathematics in Imaging}, pp.~MW2D--3, Optical Society of America, 2018.

\bibitem{jbacca}
J.~Bacca, S.~Pinilla, and H.~Arguello, ``Coded aperture design for
  super-resolution phase retrieval,'' in {\em 2019 European Association for
  Signal Processing (EUSIPCO), (accepted)}.

\bibitem{ajerez}
A.~Jerez, S.~Pinilla, H.~Garcia, and H.~Arguello, ``Target identification from
  coded diffraction patterns via template matching,'' in {\em 2019 European
  Association for Signal Processing (EUSIPCO), (accepted)}.

\bibitem{gross2017improved}
D.~Gross, F.~Krahmer, and R.~Kueng, ``Improved recovery guarantees for phase
  retrieval from coded diffraction patterns,'' {\em Applied and Computational
  Harmonic Analysis}, vol.~42, no.~1, pp.~37--64, 2017.

\bibitem{guerrero}
A.~Guerrero, S.~Pinilla, and H.~Arguello, ``Phase recovery from designed coded
  diffraction patterns in optical imaging,'' {\em IEEE Transactions on Image
  Processing (in review)}.

\bibitem{qiu2016undersampled}
T.~Qiu and D.~P. Palomar, ``Undersampled sparse phase retrieval via
  majorization--minimization,'' {\em IEEE Transactions on Signal Processing},
  vol.~65, no.~22, pp.~5957--5969, 2017.

\bibitem{jensen1987introductory}
J.~R. Jensen and K.~Lulla, ``Introductory digital image processing: a remote
  sensing perspective,'' 1987.

\end{thebibliography}


\begin{thebibliography}{10}

\bibitem{angeles2013}
M.~de~los~Ángeles García~Juanatey, J.~Hübert, A.~Tryggvason, and L.~B. Pedersen, ``Imaging the kristineberg mining area with two perpendicular magnetotelluric profiles in the skellefte ore district, northern sweden,'' {\em Geophysical Prospecting}, vol.~61, no.~1, pp.~200--219, 2013.

\bibitem{Zhdanov2011}
M.~S. Zhdanov, L.~Wan, A.~Gribenko, M.~Čuma, K.~Key, and S.~Constable, ``Large-scale 3d inversion of marine magnetotelluric data: Case study from the gemini prospect, gulf of mexico,'' {\em GEOPHYSICS}, vol.~76, no.~1, pp.~F77--F87, 2011.

\bibitem{Guo2011}
R.~Guo, S.~E. Dosso, J.~Liu, J.~Dettmer, and X.~Tong, ``Non-linearity in bayesian 1-d magnetotelluric inversion,'' {\em Geophysical Journal International}, vol.~185, pp.~663--675, 5 2011.

\bibitem{Feng2022}
D.~Feng, X.~Su, X.~Wang, S.~Ding, C.~Cao, S.~Liu, and Y.~Lei, ``Magnetotelluric regularized inversion based on the multiplier method,'' {\em Minerals}, vol.~12, no.~10, 2022.

\bibitem{jupp1977}
D.~L.~B. Jupp and K.~Vozoff, ``{Two-dimensional magnetotelluric inversion},'' {\em Geophysical Journal International}, vol.~50, pp.~333--352, 08 1977.

\bibitem{grayver2013}
A.~V. Grayver, R.~Streich, and O.~Ritter, ``{Three-dimensional parallel distributed inversion of CSEM data using a direct forward solver},'' {\em Geophysical Journal International}, vol.~193, pp.~1432--1446, 03 2013.

\bibitem{newman2000}
G.~A. Newman and D.~L. Alumbaugh, ``{Three-dimensional magnetotelluric inversion using non-linear conjugate gradients},'' {\em Geophysical Journal International}, vol.~140, pp.~410--424, 02 2000.

\bibitem{Candansayar2008}
M.~E. Candansayar, ``Two-dimensional inversion of magnetotelluric data with consecutive use of conjugate gradient and least-squares solution with singular value decomposition algorithms,'' {\em Geophysical Prospecting}, vol.~56, no.~1, pp.~141--157, 2008.

\bibitem{Kang2017}
S.~Kang, L.~J. Heagy, R.~Cockett, and D.~W. Oldenburg, ``{Exploring nonlinear inversions: A 1D magnetotelluric example},'' {\em Leading Edge}, vol.~36, pp.~696--699, 8 2017.

\bibitem{Liu2022}
W.~Liu, H.~Wang, Z.~Xi, R.~Zhang, and X.~Huang, ``Physics-driven deep learning inversion with application to magnetotelluric,'' {\em Remote Sensing 2022, Vol. 14, Page 3218}, vol.~14, p.~3218, 7 2022.

\bibitem{Liao2022}
X.~Liao, Z.~Shi, Z.~Zhang, Q.~Yan, and P.~Liu, ``2d inversion of magnetotelluric data using deep learning technology,'' {\em Acta Geophysica}, vol.~70, pp.~1047--1060, 6 2022.

\bibitem{RahmaniJevinani2024}
M.~Rahmani~Jevinani, B.~Habibian~Dehkordi, I.~J. Ferguson, and M.~H. Rohban, ``{Deep learning-based 1-D magnetotelluric inversion: performance comparison of architectures},'' {\em Earth Science Informatics}, vol.~17, pp.~1663--1677, 4 2024.

\bibitem{Deng2023a}
F.~Deng, S.~Yu, X.~Wang, and Z.~Guo, ``{Accelerating magnetotelluric forward modeling with deep learning: Conv-BiLSTM and D-LinkNet},'' {\em GEOPHYSICS}, vol.~88, pp.~E69--E77, 3 2023.

\bibitem{Rung-Arunwan2022}
T.~Rung-Arunwan, W.~Siripunvaraporn, and H.~Utada, ``{The effect of initial and prior models on phase tensor inversion of distorted magnetotelluric data},'' {\em Earth, Planets and Space}, vol.~74, p.~51, 4 2022.

\bibitem{Zhdanov2017}
M.~S. Zhdanov, {\em {Foundations of Geophysical Electromagnetic Theory and Methods}}.
\newblock Elsevier, 2017.

\bibitem{tensorflow2015-whitepaper}
M.~Abadi, A.~Agarwal, P.~Barham, E.~Brevdo, {\em et~al.}, ``{TensorFlow}: Large-scale machine learning on heterogeneous systems,'' 2015.
\newblock Software available from tensorflow.org.

\bibitem{Feng2019AnDenseNet}
X.~Feng, H.~Yao, and S.~Zhang, ``{An efficient way to refine DenseNet},'' {\em Signal, Image and Video Processing}, vol.~13, pp.~959--965, 7 2019.

\bibitem{Liu2024}
Y.~Liu, X.~Ma, L.~Wang, C.~Yin, X.~Ren, B.~Zhang, and Y.~Su, ``{A Robust Approach for Geo-electromagnetic Sounding Data Inversion using L1-norm Misfit and Adaptive Moment Estimation},'' {\em IEEE Transactions on Geoscience and Remote Sensing}, 2024.

\bibitem{pmlr-v9-glorot10a}
X.~Glorot and Y.~Bengio, ``Understanding the difficulty of training deep feedforward neural networks,'' in {\em Proceedings of the Thirteenth International Conference on Artificial Intelligence and Statistics} (Y.~W. Teh and M.~Titterington, eds.), vol.~9 of {\em Proceedings of Machine Learning Research}, (Chia Laguna Resort, Sardinia, Italy), pp.~249--256, PMLR, 13--15 May 2010.

\bibitem{Adamw}
I.~Loshchilov and F.~Hutter, ``Fixing weight decay regularization in adam,'' {\em CoRR}, vol.~abs/1711.05101, 2017.

\bibitem{Heagy2017}
L.~J. Heagy, R.~Cockett, S.~Kang, G.~K. Rosenkjaer, and D.~W. Oldenburg, ``{A framework for simulation and inversion in electromagnetics},'' {\em Computers {\&} Geosciences}, vol.~107, pp.~1--19, 10 2017.

\bibitem{Cockett2015}
R.~Cockett, S.~Kang, L.~J. Heagy, A.~Pidlisecky, and D.~W. Oldenburg, ``{SimPEG: An open source framework for simulation and gradient based parameter estimation in geophysical applications},'' {\em Computers {\&} Geosciences}, vol.~85, pp.~142--154, 12 2015.

\bibitem{USGS}
{U.S. Geological Survey (USGS)}, ``{ Data U.S. Geological Survey},'' 2022.

\bibitem{Patel2020}
P.~Patel, K.~Mohan, and P.~Chaudhary, ``{Estimation of sediment thickness (including Mesozoic) in the western central part of Kachchh Basin, Gujarat (India) using Magnetotellurics},'' {\em Journal of Applied Geophysics}, vol.~173, p.~103943, 2 2020.

\end{thebibliography}
    \bibliographystyle{ieeetr}
    
\end{document}